\documentclass[useAMS,usenatbib]{mn2e}
\usepackage{graphicx,epsfig}

\newcommand{\ud}{\mathrm{d}}

\def\kms{\mbox{\,km~s$^{-1}$}}
\def\kmsmpc{\mbox{\,km~s$^{-1}$~Mpc$^{-1}$}}
\newcommand{\lfir}{\hbox{L$_{\rm FIR}$}}

\def\cmmm{\mbox{~cm$^{-3}$}}
\def\nism{n_{\rm ISM}}
\def\Tism{T_{\rm ISM}}
\def\mdot{\dot M}
\def\msw{M_{\rm sw}}

\newcommand{\Msun}{\hbox{~${\rm M}_\odot$}}
\newcommand{\msun}{\hbox{~${\rm M}_\odot$}}
\newcommand{\ml}{~\Msun ~\rm yr^{-1}}
\newcommand\msunyr{\mbox{\,${\rm M_{\odot}\, yr^{-1}}$}}
\newcommand{\ergshz}{\mbox{~erg~s$^{-1}$~Hz$^{-1}$}}

\newcommand{\Lsun}{\thinspace\hbox{$\hbox{L}_{\odot}$}}
\def\ccsnrate{\mbox{\,$r_{\rm CCSN}$}}
\newcommand{\EE}[1]{\hbox{$10^{ #1 }$}}

\def\lsim{\!\!\!\phantom{\le}\smash{\buildrel{}\over
 {\lower2.5dd\hbox{$\buildrel{\lower2dd\hbox{$\displaystyle<$}}\over
                                 \sim$}}}\,\,}
\def\gsim{\!\!\!\phantom{\ge}\smash{\buildrel{}\over
{\lower2.5dd\hbox{$\buildrel{\lower2dd\hbox{$\displaystyle>$}}\over
                               \sim$}}}\,\,}
\def\EE#1{\times 10^{#1}}

\title[Radio monitoring of NGC 7469 and its host SN 2000ft]
{Radio monitoring of NGC 7469: 
Late time radio evolution of SN 2000ft and the circumnuclear starburst in NGC 7469
}
\author[M.~A P\'erez-Torres et al.] 
{M.A.\ P\'erez-Torres$^1$\thanks{E-mail: torres@iaa.es}, 
A.\ Alberdi$^1$, 
L. Colina$^2$,
J.M. Torrelles$^3$, 
N. Panagia$^{4,5,6}$,
\newauthor
A. Wilson$^{7\,\dagger}$, 
E. Kankare$^{8,9}$, 
S. Mattila$^8$\\
$^1$Instituto de Astrof\'{\i}sica de Andalucia, IAA-CSIC, Apdo. 3004, 18080 Granada, Spain\\
$^2$Instituto de Estructura de la Materia, IEM-CSIC,  28006 Madrid, Spain\\
$^3$Instituto de Ciencias del Espacio (CSIC)-IEEC, Facultat de F\'{\i}sica,
Universitat de Barcelona, Mart\'{\i} i Franqu\`es 1, 08028 Barcelona, Spain\\
$^4$STScI, Baltimore, MD 21218, USA\\
$^5$INAF-CT, Osservatorio Astrofisico di Catania, I-95123 Catania, Italy\\
$^6$Supernova Ltd. OYV \# 131, Virgin Gorda, British Virgin Islands\\
$^7$Astronomy Department, Univ. of Maryland, College Park, MD 20742, USA. $^\dagger$Deceased.\\
$^8$Tuorla Observatory, Department of Physics and Astronomy, University of Turku, V\"ais\"al\"antie 20, FI-21500 Piikki\"o, Finland\\
$^{9}$Nordic Optical Telescope, Apartado 474, E-38700 Santa Cruz de La Palma, Spain\\
}

\date{Accepted 2009 July 14.  Received 2009 July 13; in original form 2009 June 10}

\pagerange{\pageref{firstpage}--\pageref{lastpage}} \pubyear{2005}
\def\LaTeX{L\kern-.36em\raise.3ex\hbox{a}\kern-.15em
T\kern-.1667em\lower.7ex\hbox{E}\kern-.125emX}

\begin{document}
\label{firstpage}
\maketitle

\begin{abstract}

  We present the results of an eight-year long monitoring of the radio emission from the Luminous Infrared Galaxy (LIRG) NGC 7469, using 8.4 GHz Very Large Array (VLA) observations at 0.3'' resolution.  Our monitoring shows that the late time evolution of the radio supernova SN 2000ft follows a decline very similar to that displayed at earlier times of its optically thin phase.  The late time radio emission of SN 2000ft is therefore still being powered by its interaction with the presupernova stellar wind, and not with the interstellar medium (ISM).  Indeed, the ram pressure of the presupernova wind is $\rho_w\,v_w^{2} \approx 7.6\EE{-9}$\,${\rm dyn}\,{\rm cm}^{-2}$, at a supernova age of $t \approx 2127$ days, which is significantly larger than the expected pressure of the ISM around SN 2000ft.  At this age, the SN shock has reached a distance $r_{sh}\approx 0.06$ pc, and our observations are probing the interaction of the SN with dense material that was ejected by the presupernova star about 5820 years prior to its explosion.  From our VLA monitoring, we estimate that the swept-up mass by the supernova shock after about six years of expansion is $\msw \approx 0.29$\msun, assuming an average expansion speed of the supernova of 10$^4$\kms.

  We also searched for recently exploded core-collapse supernovae in our VLA images.  Apart from SN 2000ft ($S_\nu \approx 1760 \mu$Jy at its peak, corresponding to $1.1\EE{28}$\ergshz), we found no evidence for any other radio supernova (RSN) more luminous than $\approx 6.0\EE{26}$\ergshz, which suggests that no other Type IIn SN has exploded since 2000 in the circumnuclear starburst of NGC 7469.

\end{abstract}


\section{Introduction}
\label{intro}

The central kiloparsec region of many nearby Luminous Infra-Red Galaxies (LIRGs) shows a distribution of their radio emission consisting of a compact ($\leq$ 150 pc), high surface brightness, central radio source immersed in a low surface brightness circumnuclear halo (Condon et al. 1991).  While the compact, centrally located radio emission in these galaxies might be generated by a point-like source (AGN), or by the combined effect of multiple RSNe and supernova remnants (SNRs), e.g., the evolved nuclear starburst in Arp~220 (Parra et al. 2007 and references therein) and M~82 (e.g. Muxlow et al. 1994), it seems now well established that in the circumnuclear regions of those objects there is an ongoing burst of star formation producing core-collapse supernovae (CCSNe, Type Ib/c and Type II SNe) at a high rate (e.g.  NGC 7469, Colina et al. 2001a,b; Arp 299, Neff et al. 2004; NGC 6240, Gallimore \& Beswick 2004; Arp 220, Parra et al. 2007; IRAS~18293-3413, Mattila et al. 2007, P\'erez-Torres et al. 2007; IRAS 17138-1017, Kankare et al. 2008a, P\'erez-Torres et al. 2008, Kankare et al. 2008b).  Given a reasonable assumption for the initial mass function (IMF) (e.g., Colina \& P\'erez-Olea 1992; Smith et al. 1998), the detection of {CCSNe} can be used to characterize the most important parameter in starbursts, namely the star formation rate (SFR).  However, the estimation of the CCSN rate in LIRGs is a challenging task, as the optical emission of supernovae is hampered by background brightness and by large amounts of dust in the nuclear starburst environment, and therefore those SNe remain undiscovered by optical searches.

Fortunately, it is possible to directly probe the recent star forming activity in LIRGs by means of high-resolution, high-sensitivity radio searches for CCSNe, since these wavelengths do not suffer from dust obscuration, and can therefore set strong constraints on the properties of star formation in the dust-enshrouded environments encountered in LIRGs.  Significant radio emission from CCSNe is expected, as the interaction of the SN ejecta with the circumstellar medium (CSM) gives rise to a high-energy density shell, which is Rayleigh-Taylor unstable and drives turbulent motions that amplify the existing magnetic field in the presupernova wind, and efficiently accelerate relativistic electrons, thus enhancing the emission of synchrotron radiation at radio wavelengths (Chevalier 1982).  Therefore, starburst activity in the circumnuclear regions of LIRGs ensures both the presence of a high number of massive stars and a dense surrounding medium, so bright radio SNe are expected to occur (Chevalier 1982, Chugai 1997).  In fact, radio and optical/IR observations have shown that highly-extinguished SNe do exist in the circumnuclear ($r\lsim$1~kpc) region of local LIRGs (e.g., SN 2000ft in NGC 7469, Colina et al. 2001a,b; SN 2004ip in IRAS 18293-3413, Mattila et al. 2007, P\'erez-Torres et al. 2007; SN 2008cs in IRAS 17138-1017, Kankare et al. 2008a, P\'erez-Torres et al. 2008, Kankare et al. 2008b).

NGC 7469 is a well known barred spiral galaxy located at a distance of 70 Mpc (Sanders et al. 2003; $H_0$= 70 \kmsmpc; at the distance of NGC 7469, 1 arcsec corresponds to a linear size of 333 pc) containing a luminous, QSO-like, Seyfert 1 nucleus surrounded by a dusty circumnuclear starburst of 5'' (1.6 kpc) in diameter (Wilson et al. 1991, Malkan et al. 1998, Genzel et al. 1995, Scoville et al. 2000, D\'{\i}az-Santos et al. 2007). NGC 7469 is also a LIRG with L$_{\rm IR}$(8-1000$\mu$m) = 4.5$\times 10^{11} \Lsun$ (obtained from Sanders et al. 2003 for our adopted distance to NGC 7469).  During the course of a Very Large Array (VLA) monitoring program aimed at the detection of RSNe in NGC 7469, Colina et al. (2001a) detected a strong compact radio source in the circumnuclear starburst of this galaxy, which was later identified with a RSN (Colina et al. 2001b).  The RSN was assigned the name SN 2000ft (Colina et al. 2002), and was the first RSN discovered in the circumnuclear ring of a Seyfert~1, Luminous Infrared Galaxy, and its detection confirmed that radio observations are a powerful tool for detecting RSNe up to large distances in the local Universe.  SN 2000ft is located at a projected distance of 600 pc from the QSO-like nucleus of NGC 7469, and showed an 8.4 GHz emission peak of 1.1 $\times 10^{28}$ \ergshz.  The case of SN~2000ft is outstanding, since its high brightness has allowed a multi-frequency, long term monitoring of its radio emission since its discovery.  Alberdi et al. (2006) found, based on the flux density evolution of SN 2000ft for the first $\lsim$1000 days after its explosion, that SN 2000ft shares many of the same properties that are common to RSNe optically identified as Type II SNe, despite having exploded in the dusty and very dense environment of a LIRG circumnuclear region.  Finally, we also note that the optical glow of SN 2000ft has recently been identified in archival HST images taken on 13 May 2000 (Colina et al. 2007), consistent with the predicted date of the explosion from the analysis of its radio light curves.

\begin{figure*}
{\includegraphics[totalheight=0.8\textheight]{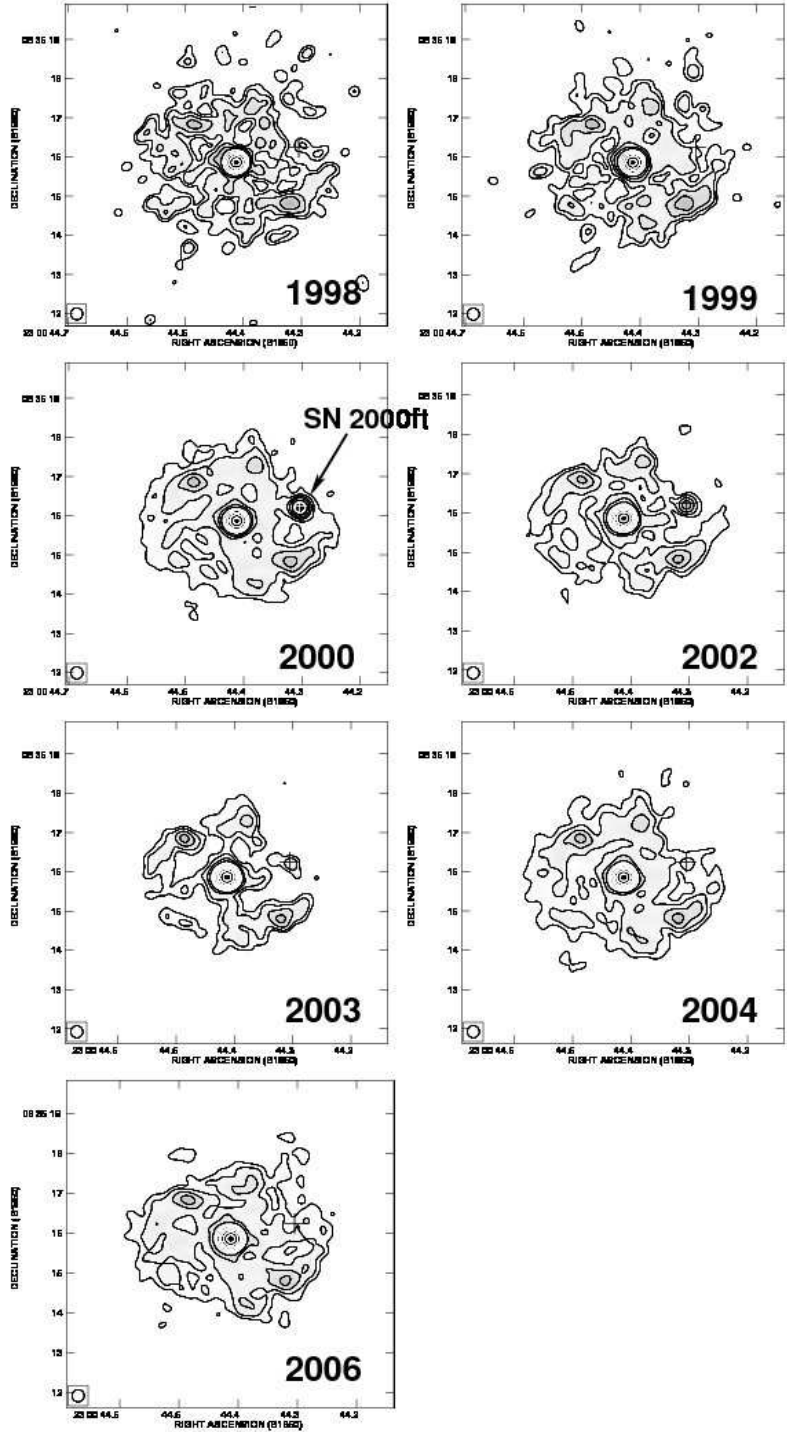}}
\caption[]{Total intensity radio images of NGC~7469 obtained at 8.4~GHz
  with the VLA in A-configuration, which cover a time baseline of
  almost eight years (from 8 April 1998 till 5 February 2006). The
  maps have been spatially convolved to obtain the same circular
  Gaussian beam of size 0.31'', to be able to compare adequately all
  the images.  The contour levels are drawn at
  (3,$3\,\sqrt{3}$,9,...)$\times 25\mu$ Jy beam$^{-1}$, which is the
  highest off-source rms of all the maps (corresponding to the 1998
  and 1999 epochs).  The peaks of brightness are of 12.61, 12.40,
  12.27, 12.53, 13.01, 12.91, and 13.15~mJy beam$^{-1}$ for the 1998
  through 2006 epochs, respectively, and correspond to the nucleus of
  NGC~7469. The cross corresponds to the position of the peak of SN 2000ft
  in the 2000 image. }
\label{fig1}
\end{figure*}

\section{VLA Observations and Data Analysis}

We have monitored the radio emission from the LIRG NGC~7469 since
1998, using the VLA\footnote{The National Radio Astronomy Observatory is a facility of the National
  Science Foundation operated under cooperative agreement by
  Associated Universities Inc.} in A configuration at 8.4 GHz
(FWHM$\simeq$0.23-0.31 arcsec).  We observed NGC~7469 on 8 April 1998,
8 September 1999, 27 October 2000, 8 February 2002, and 19 June 2003
(see Colina et al. 2001b and Alberdi et al. 2006), and on 11 November
2004 and 5 February 2006 (this paper), using an effective bandwidth of
100~MHz and both circular polarizations.  Each observing run lasted
about 4~hr (except those on 8 April 1998 and 8 September 1999, which
lasted about 50 and 40 minutes each, with effective integration
on-target times of $\sim$40 and $\sim$30 minutes).  A typical run
consisted of $\sim$16~min scans on the nucleus of NGC~7469,
interleaved with $\sim$2.5~min scans on the phase and amplitude
calibrator 2251+158, and each time ending with a $\sim$6~min observation
of the quasar 3C~48 to set the absolute flux density scale. This
observing scheme resulted in effective integration times on NGC 7469
of $\approx$3.0 hr.  We later edited, calibrated, and imaged the VLA
data corresponding to all epochs by following standard data reduction
techniques implemented within the NRAO Astronomical Imaging Processing
System ({\it AIPS}).

\begin{figure}
\begin{center}
\resizebox{8.6 cm}{!}{\includegraphics[width=8.5cm]{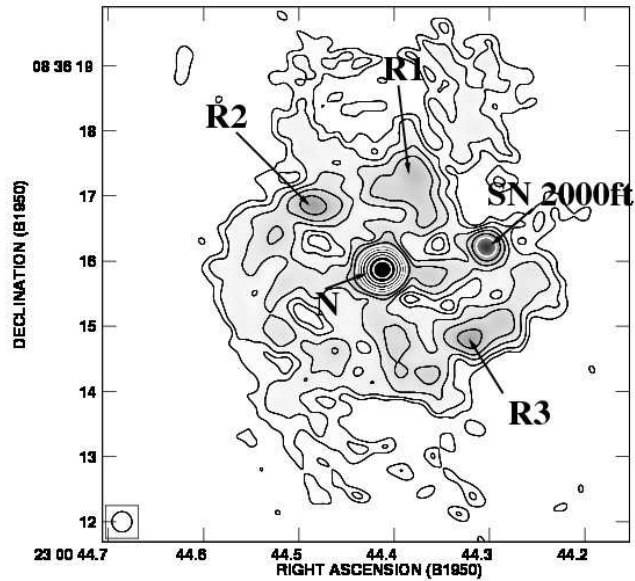}}
\caption[]{Total intensity radio image of NGC~7469 obtained at 8.4~GHz
  with the VLA in A-configuration, using all $u-v$ data for epochs
  1998 through 2006, whose individual images are shown in Figure
  \ref{fig1}.  The radio image has been obtained by convolving the
  data with a circular Gaussian beam of size 0.31''.  The contour
  levels are drawn at (3,$3\,\sqrt{3}$,9,...)$\times$ the off-source
  rms of the image, which is 10\,$\mu$Jy beam$^{-1}$. This is the
  deepest 8.4 GHz image of NGC 7469 ever produced. We indicate the
  positions of the nucleus (N),  the RSN SN
  2000ft, and the star forming regions R1, R2, and R3.}
\label{n7469-dbcon}
\end{center}
\end{figure}

The radio maps shown in Fig.~\ref{fig1} were obtained after a few
iterations of imaging and phase-selfcalibration of the NGC~7469 data,
setting ROBUST=2 in task IMAGR, and the achieved synthesized beam
ranged between 0.26'' and 0.31'', depending on the u-v coverage at a
given observing epoch.  To facilitate comparisons among the observing
epochs, the maps have been spatially convolved to obtain a same
circular Gaussian beam of size 0.31'' (see Fig.~\ref{fig1}), which
corresponds to the lowest resolution image on 8 April 1998.  The
off-source rms for each map is of 25, 25, 15, 17, 20, 14, and
14\,$\mu$Jy beam$^{-1}$ for the 1998, 1999, 2000, 2002, 2003, 2004,
and 2006 images, respectively, which are in general close to the
theoretically expected image thermal noise values.  Figure
\ref{n7469-dbcon} shows the resulting image of combining the radio
interferometric data for all epochs (1998 through 2006), where we have
marked for convenience the positions of the nucleus (N), the
well-known supernova SN 2000ft, and the star forming regions R1, R2,
and R3 (Colina et al. 2001b).

\section{Results}
\label{results}

\subsection{Radio emission from NGC 7469 and its SN 2000ft}
\label{res-late-time}

We summarize the results of our eight-year long VLA-A monitoring of NGC
7469 and its radio bright SN 2000ft in Table \ref{tab:vla-log} and
Figure \ref{fig1}.  The average measured 8.4 GHz total flux density of
the galaxy is $\sim$35 mJy.  A significant fraction of this radio
emission ($\sim$15-17 mJy, depending on the epoch), comes from the
nuclear region of the galaxy, with an angular size of $\approx$0.5''
($\approx$167 pc), which hosts an AGN.  We also note that our MERLIN
(Multi-Element Radio Linked Interferometer Network) observations at 5
GHz (Alberdi et al. 2006), imply a linear size $\lsim$50 pc for the
nuclear region.  Further, recent high-resolution images of this
region obtained with the European VLBI Network (EVN) show that the
putative AGN consists of at least four compact components (Alberdi et
al., in preparation).

\begin{table*}
\caption[]{8.4 GHz positions and flux densities of compact sources in 
the (circum)nuclear regions of NGC 7469.
The uncertainty in the flux density determination results from adding
in quadrature the off-source rms of each map and a 2\% of the local
maximum and, linearly, the standard deviation of the locally
subtracted background emission.}
\label{tab:vla-log}

\tabcolsep 5pt
\begin{tabular}{lcclllllll}
\hline
Source & $\alpha$ & $\delta$ & \multicolumn{7}{c}{S$_\nu$ ($\mu$Jy)} \\\cline{4-10}
     & (B1950)  & (B1950)  & 8 Apr 1998 & 8 Sep 1999 & 27 Oct 2000 & 8 Feb 2002 & 19 Jun 2003 & 11 Nov 2004 & 5 Feb 2006 \\
\hline \hline
SN 2000ft & 23 00 44.305 & 08 36 16.24& 
          $\lsim$28 & $\lsim$29 & 1762$\pm$51 & 412$\pm$32 & 116$\pm$33
          & 75$\pm$27 & 36$\pm$27 \\
Nucleus  & 23 00 44.412 & 08 36 15.86 
         & 12610$\pm$266 & 12400$\pm$262 & 12270$\pm$259 & 12530$\pm$264
         & 13010$\pm$274 & 12910$\pm$272 & 13150$\pm$276\\
R1       & 23 00 44.377 & 08 36 17.23
         & 340$\pm$39          &  350$\pm$39           &  330$\pm$29          & 300$\pm$31
         & 280$\pm$34          &  310$\pm$28           & 280$\pm$28\\
R2       & 23 00 44.485 & 08 36 16.84
         & 460$\pm$39           &  510$\pm$40          &  470$\pm$31          & 430$\pm$32
         & 450$\pm$35           &  460$\pm$28          &  480$\pm$30\\
R3       & 23 00 44.341 & 08 36 14.59  
         & 550$\pm$40          &  500$\pm$40          &  470$\pm$31         & 440$\pm$32
         & 430$\pm$35          &  450$\pm$30          &  440$\pm$30\\
\hline
\end{tabular}
\end{table*}

Our 8.4 GHz VLA images clearly show the appearance of a radio supernova, SN 2000ft, whose extreme brightness and long-lasting behavior confirmed not only its core-collapse nature, but also that it was a Type IIn supernova (Alberdi et al. 2006). The progenitors of those SNe are thought to be rather massive stars, about (18$-$30)$\msun$ (Weiler et al. 1990, van Dyk et al. 1993), although we should note that there is also recent evidence that some of the Type IIn progenitors might be much more massive, e.g. Gal-Yam et al. (2007), Trundle et al. (2008).  We obtained the flux density of SN 2000ft for each image in Figure \ref{fig1}, after subtraction of the background emission, i.e., the local, secular, non variable emission of the galaxy at 8.4 GHz.  We estimated the local background radio emission in the region around SN 2000ft by four different procedures: (i) We used the {\it AIPS} task IMFIT, solving for a zero level component (the background) for each observing epoch. This procedure yielded background values of 123, 96, 88, 100, and 99 microJy for the 2000, 2002, 2003, 2004, and 2006 epochs, respectively.  The above background values were subtracted from the peak values estimated for SN 2000ft, and are the values quoted in Table~\ref{tab:vla-log}.  (ii) We also determined the background emission at the position of the supernova as the difference in the flux density of two images, the first one obtained by using a purely naturally weighting scheme (UVWTFN='NA'; ROBUST=5), and the second one obtained by using a uniform weighting scheme (UVWTFN=' '; ROBUST=-5), where in addition we dropped the inner 35 kilo-$\lambda$ of the u-v range, essentially removing all extended emission from the data.  The former image yields the lowest off-source r.m.s. in the images, and is most sensitive to the extended, secular background emission, while the second is essentially free of background emission, at the expense of having a higher off-source rms. The estimates for the secular background emission obtained in this way coincided with those obtained by following the first procedure within 2 $\mu$Jy/b.  As a measure of the uncertainty in the locally subtracted background emission we took the standard deviation of the above values (13$\mu$Jy/b). We conservatively added this uncertainty linearly, rather than in quadrature, to the total uncertainty in the flux density measurements for SN 2000ft, which are listed in Table \ref{tab:vla-log} and plotted in Figure \ref{fig,lcurve}. Those values show that the late time radio emission of SN 2000ft is well characterized by a single index for the power-law time decay, even six years after its explosion ($S_\nu \propto t^{-2.02^{+0.14}_{-0.07}}$; see Sect. \ref{sec,lcurve} for details).

We note here that the flux density values reported in Table \ref{tab:vla-log} for SN 2000ft are somewhat different from those presented in Alberdi et al. (2006).  Their values were determined from images with different synthesized beams, used different weighting imaging schemes, applied amplitude selfcal after a few iterations of phase-selfcalibration, and used different field and cell sizes for each of the images presented in their paper.  Here, we followed a homogeneous data reduction scheme. Specifically, we used the same field and cell sizes for all images, made a standard imaging scheme, using natural weighting and applying phase-selfcalibration only.  The final run of IMAGR was done using a circular Gaussian beam of size 0.31 arcsec for all epochs. Finally, we took into account the subtraction of the background radio emission. All this explains the differences in the flux densities reported in those papers \footnote{The only exception is the flux density reported for
  the discovery of SN 2000ft, which is mistakenly given as 1.60 mJy in
  Alberdi et al. (2006), while the value that should have been
   reported is 1.70 mJy.}, but do not affect, or modify, any of the
results or conclusions obtained in either paper.

\begin{figure}
\begin{center}
\resizebox{8.6 cm}{!}{\includegraphics[width=8.5cm,angle=-90]{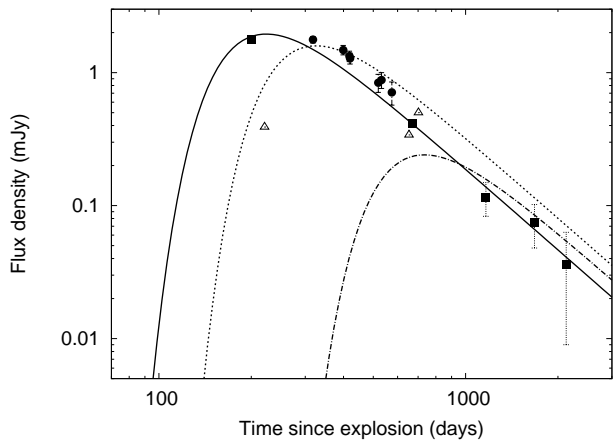}}
\caption[]{Radio light curve of SN 2000ft at 8.4 GHz (filled squares), 5.0
  GHz (filled circles), and 1.6 GHz (upper limits; open triangles),
  along with the resulting light curve fits (8.4 GHz, solid line; 5.0
  GHz, dotted line; 1.4 GHz, dashed-dotted line) of the parameterized
  model described in Section \ref{sec,lcurve}. The explosion date is
  fixed to 10 May 2000, three days before its serendipitous optical
  discovery (Colina et al. 2007). The 8.4 GHz data are from this work,
  while data at other wavelengths were taken from Alberdi et
  al. (2006).  Note that the late time data ($t \gsim 1000$ days)
  follows genuinely the same steep, power-law decline found at earlier
  times.}
\label{fig,lcurve}
\end{center}
\end{figure}

\subsection{Radio light curve of SN 2000ft}
\label{sec,lcurve}

We show in Figure \ref{fig,lcurve} the flux density evolution of SN
2000ft. The data plotted correspond to our new flux density estimates
(corrected for the galaxy background emission), as well as our late
time ($t \gsim 1000$ days) radio data at 8.4 GHz (see Table
\ref{tab:vla-log}), while data at other wavelengths are taken from
Alberdi et al. (2006). (Note that since the data are other frequencies
were obtained with MERLIN, which has a resolution of $\sim$40 mas at 5
GHz, the putative background radio emission was resolved
out). Following Alberdi et al. (2006), we fitted the light curves in
terms of the ``mini-shell'' model (Chevalier 1982), including
modifications by Weiler et al. (1990).  In particular, Alberdi et
al. (2006) found that both internal opacity (consisting of synchrotron
self-absorption and mixed thermal free-free absorption/nonthermal
absorption) and clumpy opacity contributions were not required to fit
the data. Alberdi et al. (2006) did find, however, that a distant
foreground absorber was needed to account for the non-detection of SN
2000ft at 1.6 GHz in any of the observing epochs, so we have also
fitted for such an absorber.  In addition, we also use the fact that
SN 2000ft has been recently detected at optical wavelengths on 13 May
2000 (Colina et al. 2007), based on an analysis of HST archival data.
We have fixed here the explosion date to 10 May 2000, so as to make it
compatible with the optical detection. Therefore, we fitted our
multi-wavelength observations using the following expression (e.g. Weiler et al. 2002):

\begin{equation}
S_\nu = K_1\, \nu^\alpha_5\, t_1^\beta\, e^{-\tau_{ext}}
\label{eq,fit}
\end{equation}

\noindent 
where $S_\nu$ is the observed flux density, in mJy; $\nu_5$ is the
observing frequency, in units of 5 GHz; $t_1$ is the time since the
explosion date, in days; and $\tau_{ext}$ is the external optical
depth, assumed to arise from purely thermal, ionized hydrogen that
absorb with frequency dependence $\nu^{-2.1}$.  In turn, $\tau_{ext} =
\tau_{CSM}+ \tau_{dist}$, $\tau_{CSM} =
K_{2}\,\nu_5^{-2.1}\,t_1^{\delta}$, $\tau_{dist} =
K_{4}\,\nu_5^{-2.1}$, where $K_1$, $K_2$, and $K_4$ correspond to the
flux density ($K_1$) and uniform optical depths ($K_2$, $K_4$) at 5
GHz, one day after the explosion, respectively.  The parameter
$\delta$ accounts for the time dependence of the optical depth for the
local uniform media.

Our best fit is yielded by the following set of parameters: spectral index, $\alpha=-1.27$; power-law time decay, $\beta=-2.02$; $K_{1}= 4.45 \times 10^{5}$ mJy; $K_{2}= 1.67 \times 10^{7}$; adopted power-law time decay of the optical depth, $\delta=-2.94$; foreground uniform optical depth $\tau_{ff}$ (most likely associated with an H II region, as found by Alberdi et al. 2006), $K_4 \gsim$0.17.

\subsection{A systematic search for radio supernovae in NGC 7469}
\label{search}

RSN SN~2000ft was extraordinarily bright, and therefore was easy to detect and confirm as such in our images. However, RSNe (RSNe) are on average intrinsically fainter than SN 2000ft, and hence several RSNe could have exploded since 1998 but gone unnoticed by simple visual inspection of our VLA radio images of NGC 7469, especially since there is a non-negligible amount ($\sim 100 \mu$Jy) of secular, non-variable, galactic radio emission.  We therefore reverted to the use of other methods that could hint to the explosion of dimmer supernovae during this period.

In particular, for the purpose of searching for RSNe fainter than SN 2000ft, we used all of our seven VLA-A observing epochs, including two epochs prior to the discovery of SN 2000ft, and re-analyzed them in a similar, homogeneous manner, including editing, calibrating, and imaging the data following standard procedures within {\it AIPS}.
We then used a slightly modified version (see Melinder et al. 2008) of the Optimal Image Subtraction (Alard \& Lupton 1998; Alard 2000) package ISIS 2.2. The method matches the PSF, intensities, and background of the better ``seeing'' image, to the aligned poorer ``seeing'' images from the different epochs before the subtraction. The method has been used successfully to detect supernovae within the highly obscured nuclear regions of luminous infrared and starburst galaxies in near-IR, e.g. the discovery of SN 2004ip (Mattila et al. 2007). However, to our knowledge this is the first time it is applied to search for supernovae in radio images.  The radio nucleus of the host galaxy was selected as the image region to derive the convolution kernel. The epoch from 2004 was used as the reference frame.
No significant effect on the image subtraction quality was found when
changing the kernel basis function width parameters. Flux measurements
were performed on the subtracted images using {\sc IRAF} and {\sc
  GAIA} software packages. Due to the incomplete (u-v) coverage and the use of the CLEAN deconvolution algorithm, the noise pattern of the radio images is complex and the background is not smooth in the subtracted images and contains a large number of false positive and negative ``sources''. Therefore, to determine our SN detection threshold we measured the flux in 44 circular apertures around the nucleus in the subtracted images using an aperture of 60 pixels radius and a sky annulus with an inner and outer radius of 90 and 120 pixels, respectively (the pixel size was of 0.008 arcsec). The standard deviation of these measured values was used as a $1\sigma$ limit, and corresponded to 20$\mu$Jy/b. No additional transient sources besides SN 2000ft above a $3\sigma$ detection threshold were apparent in the data set. We repeated the above procedure, taking each time a different reference epoch, but the results did not change: no additional RSNe were detected.

We also searched for intrinsic variability in regions R1, R2, R3, and N.  Since regions R1, R2, and R3 are very bright and show spectral indices typical of star forming regions (Alberdi et al. 2006), significant variability in their flux density well above statistical fluctuations must be due to an intrinsic variation caused by, most likely, a supernova.  We therefore extracted the 8.4 GHz peaks of brightness from the Sy 1 nucleus of the galaxy (N), and from the star forming regions R1, R2, and R3.  Those values are also listed in Table \ref{tab:vla-log}. However, given the uncertainties in the total flux density values of those regions, the variations are compatible with purely statistical fluctuations. Hence, there is no evidence of significant variability that could point towards a supernova.

\section{Discussion}
\label{discussion}

The main goal of our VLA observations was twofold.  First, we aimed at monitoring the radio evolution of SN 2000ft, especially at late times, and we discuss in detail this aspect in Section \ref{discussion-late-time}. Second, we searched for recently exploded core-collapse supernovae (CCSNe, i.e., Type Ib/c and Type II) in the circumnuclear starburst of NGC~7469, aimed at establishing its CCSN rate, independently of models.  Since we obtained in general very high-sensitivity (off-source rms $\lsim 25 \mu$Jy/beam) radio images of NGC 7469 at high-resolution (FWHM$\approx$0.30''), this could have permitted the discovery of new RSNe in the eight years covered by our observations. We showed in Section \ref{results} that, apart from SN 2000ft, there is no evidence for any other bright RSN in the (circum)nuclear region of NGC 7469 since 1998.  We discuss the implications of these results in terms of CCSN rates and starburst models in Section \ref{ccsnrate}.

\subsection{The mass-loss rate of SN 2000ft}
\label{discussion-lcurve}

The new fit to the SN 2000ft light curve (see Section \ref{sec,lcurve} and Figure \ref{fig,lcurve}) shows that the best fit parameters are somewhat different from those published in Alberdi et al. (2006), due to the addition of the new data at 8.4 GHz and also because the optical detection of SN 2000ft on 13 May 2000 (Colina et al. 2007) implies that the explosion happened before, even if close to, the date of the optical detection.  While the resulting fits do not change any of the facts described in Alberdi et al. (2006), one relevant parameter for the interaction of SN 2000ft with its surroundings deserves particular attention: its mass-loss rate.

In the standard circumstellar interaction model for RSNe, the synchrotron radio emission from the supernova is partially free-free absorbed by ionized gas outside the forward supernova shock front.  In this scenario, the supernova goes from an optically thick regime to an optically thin one when the peak optical depth equals $\tau_\nu^{ff} = \beta / \delta$ (this can be found by computing the time derivative in Equation \ref{eq,fit}).  In our case, we have $\tau_\nu^{ff}$(peak)=0.69. This optical depth is attained in the radio regime at an age

\begin{equation}
t_{ff}\approx 326
\left(\dot M_{-4}\over v_{w1}\right)^{2/3}\,T_{cs2}^{-1/2}\,V_4^{-1}
\,\nu_{8.45}^{-2/3}{\rm~days},
\label{eq,tau_ff}
\end{equation}

\noindent
where $\dot M_{-4}$ is the mass-loss rate in units of $10^{-4}\ml$,
$T_{cs2}$ is the circumstellar temperature in units of $2\times 10^4$ K, $V_4$
is the supernova shock velocity in units of $10^4$\kms, $v_{w1}$ is
the presupernova wind velocity in units of $10\kms$, and $\nu_{8.45}$
is the observing frequency in units of 8.45 GHz..  Since we are
interested in the mass-loss rate, we can write the previous equation
as follows:

\begin{equation}
\dot M \approx 1.66\EE{-5}\, t^{3/2}_{100}\, T_{cs2}^{3/4} V_4^{3/2} v_{w1} \nu_{8.45}
\ml
\label{eq,mass-loss}
\end{equation}

\noindent
where $t_{100}$ is the time to reach optical depth unity to free-free
absorption at a given frequency, in units of 100 days.  Therefore, we
can use our fitted light curves to directly measure the time at which
SN 2000ft reached $\tau_\nu^{ff} \approx 0.69$, which corresponds to
the emission peak at a given frequency. In particular, we find that
$\tau^{ff} \approx 0.69$ at $\approx$200 days and $\approx$300 days at
8.45 and 5.0 GHz, respectively. Substituting those values in Equation
\ref{eq,mass-loss}, we find that $\dot M \lsim (4.7-5.1)\EE{-5}
\ml$, typical of a red supergiant progenitor star, and where the
inequality takes into account that, if synchrotron-self absorption
plays some role around the observed peak of emission, then the
obtained mass-loss rate is actually an upper limit.

\subsection{Interaction around SN 2000ft}
\label{discussion-late-time}

It has been proposed that CCSNe exploding in dense environments, like
those encountered in the circumnuclear regions of starburst galaxies,
can result in very luminous RSNe (Chevalier 1982, Chugai
1997). An apparent confirmation of such a theory came from the
discovery of the radio bright SN 2000ft (Colina et al. 2001a,b) in the
circumnuclear starburst ring ($\sim$600 pc from the nucleus) of NGC
7469.  However, the VLA radio monitoring of SN 2000ft for the first
$\sim$1100 days after its explosion showed that this RSN
shared essentially the same properties that are common to radio SNe
identified as Type II SNe (Weiler et al. 2002), despite having
exploded in the dusty and very dense environment of the circumnuclear
region of NGC 7469. Indeed, Colina et al. (2007) have recently
identified the optical glow of SN 2000ft using archival HST images
taken on 13 May 2000, and note that the supernova appears to have
exploded in a region of increased local extinction ($\sim$4.2
magnitudes) near the edge of a strong lane of dust surrounding the
ring (see their color map in Figure 1), which confirms that SN 2000ft
exploded in a dense and dusty surrounding medium.

Alberdi et al. (2006) also predicted that once the interaction of the
ejecta of SN 2000ft with the interstellar medium (ISM) would become
relevant, the flux density decay time scale for SN 2000ft should be
longer than for normal luminous RSNe, whose radio emission continue
to fall off very quickly.  Such change in the flux density decaying
rate of the supernova would indicate the termination of the
presupernova stellar wind, where the pressure of the ISM is
approximately equal to the ram pressure of the wind, and represents
the passage of supernova to supernova remnant phase. This passage is
actually expected to happen earlier for supernovae exploding in dense
environments--as those encountered in starburst galaxies--than in the
less dense environments of normal galaxies.  In fact, the duration of
the radio SN phase (when the radio emission is governed by interaction
between the ejecta and its circumstellar medium, CSM) is limited by
the extent of the expanding presupernova wind, since the latter
eventually reaches a radius where its ram pressure, $\rho_w\,v_w^{2}$,
equals the external pressure of the interstellar medium (ISM),
$P_{ISM}$ (Chevalier \& Fransson 2001).  For a spherically symmetric,
steady wind ($\rho_w \propto r^{-2}$), this radius is

\begin{equation}
r_w \approx 0.18\,\dot
M_{-4}^{1/2}v_{w1}^{1/2} p_7^{-1/2}\, {\rm pc}
\label{eq,csm-ism}
\end{equation}

where $p_7 = P_{ISM}/k$ is the ISM pressure in units of $10^7$
cm$^{-3}$ K, which is the estimated pressure for the central region of
the starburst in M~82 (Chevalier \& Clegg 1985).  Once it has reached, 
a core-collapse supernova enters its supernova remnant (SNR)
phase, and its radio emission is no longer powered by the interaction
with the CSM, but rather with the ISM. At this stage, the density
encountered by the SN would be constant, which would result in a
significant flattening of the radio emission, i.e., a departure from
the general, steep decline observed in the early times of the
optically thin phase of SN 2000ft.

\subsubsection{SN 2000ft: Late radio emission still powered by interaction
with the progenitor's stellar wind}

We followed up the radio evolution of SN 2000ft using the VLA at 8.4
GHz, searching for a potential departure of the flux density decaying
rate, which would signal the passage of SN to SNR phase for SN 2000ft.
Our observations of the late time ($t \geq 1000$ days) evolution of
the 8.4 GHz emission of SN 2000ft (see Figure \ref{fig,lcurve}) do not
show evidence for any departure from the general trend observed for
the 8.4 GHz emission during its first 1000 days, and implies that the
radio emission of SN 2000ft is, after more than 2100 days, still being
powered by the interaction of the supernova with the stellar wind of
its progenitor star. It then follows that the external pressure of the
ISM is smaller than the ram pressure of the presupernova wind

\begin{eqnarray}
P_{ISM} \lsim \rho_w\,v_w^{2} & = & \dot M\,v_w/(4\,\pi\,r_w^2) \nonumber\\
    & \approx & 5.04\EE{-8}\,\dot M_{-4}\,v_{w1}\,r_{17}^{-2}\, {\rm dyn}\,{\rm cm}^{-2}
\label{eq,p_ism}
\end{eqnarray}

The supernova shock has penetrated into the presupernova wind up to a
distance $r_{sh} \approx 1.84\EE{17}$ cm ($\approx 0.06$ pc) after
2147 days of (assumed) free expansion at a constant velocity of 10000
\kms. (We should notice here that the radius may be uncertain by as
much as a factor of two, thus making $P_{ISM}$ uncertain by up to a
factor of four.)  Substituting this value, and our estimated mass-loss
rate of $\mdot \simeq 4.9 \EE{-5}$\msunyr\/ for the progenitor star of
SN 2000ft in Equation \ref{eq,p_ism}, we then obtain that the ram
pressure of the wind is $(\rho_w\,v_w^{2}) = 7.6\EE{-9}$\,${\rm
  dyn}\,{\rm cm}^{-2}$, still very high to be overcome by
$P_{ISM}$. For comparison, $P_{ISM}\approx 4\EE{-9}\, {\rm dyn}\,{\rm
  cm}^{-2}$ in the central H~II regions of M 82, and drops below
$\approx 2.1\EE{-10} {\rm dyn}\,{\rm cm}^{-2}$ at a distance of 540 pc
(Chevalier \& Clegg 1985).  We note here that Parra et al. (2007) have
reported that some of the most long-lived remnants in Arp 220 show a
rather slow, or even unnoticeable flux density decay at 18 cm, as also
found by Rovilos et al. (2005), and that this implies $P_{ISM} >
7.6\EE{-9}\, {\rm dyn}\,{\rm cm}^{-2}$. We also note that if the
environment where SN 2000ft exploded had a similar, or significantly
larger, pressure than that of the presupernova wind, we should have
expected to see a significant flattening in the flux density time
decay of SN 2000ft, which is not observed.

The number density of the thermal electrons in the stellar wind at a
distance reached by the outer shock can be written as 

\begin{equation}
n_w = 3.52\EE{4}\,M_{-4}\,v^{-1}_{w1}\,r_{17}^{-2}\, {\rm cm}^{-3}
\end{equation}

where $r_{17} = r_{sh}/10^{17}$\, cm, and we have assumed solar
abundances (i.e., the mean molecular weight of the plasma, $\mu
=0.86$). Therefore, we obtain $n_w \simeq 5.1\EE{3}$ cm$^{-3}$ at an
age of $t=2127$ days.  We stress here that the relevant physical
parameter to determine whether the SN has started to enter its SNR
phase is not the particle density ratio ($n_w/n_{\rm ism}$), but the
ratio of the ram pressure of the wind to that of the ISM.  Indeed, as
we have just shown, the thermal particle density in the preshocked
wind can be similar to that encountered in the dusty, dense
environments of LIRGS, yet the radio emission from the supernova be
powered by its interaction with the CSM.  Since we have found that
$P_{ISM}/k \lsim 5.5\EE{7}$ cm$^{-3}$\,K, one could have a rather
dense ISM ($\nism \approx 10^4$ \cmmm) at temperatures $\Tism \approx
10^3$ K, so that this gas would not be able to counterbalance the ram
pressure of the progenitor wind at the current SN age.  In this sense,
although the conditions of the surrounding ISM could be similar to
those encountered in the central regions of M~82, there is no
significant impact on both the radio emission and expansion of SN
2000ft, since the ram pressure of the presupernova wind is still
larger than that of the ISM.  In order for the ISM to affect
noticeably the radio emission and expansion of the supernova, and
assuming the above values of $\nism$ and $\Tism$, SN 2000ft will have
to expand inside its surrounding ISM up to a radius given by Equation
\ref{eq,csm-ism}, which corresponds to $r_{\rm sh} \simeq 4.3\EE{17}$
cm (= 0.14 pc), which the supernova will attain at an age of about
13.6 yr.  The SN is therefore expected to follow a similar trend to
that observed for its first 5.9 years of expansion until that age, at
which time it is expected that the SN radio emission will start to be
dominated by its interaction with the ISM, and a significant
flattening of the flux should be observed.

\subsubsection{Progenitor mass-loss history and swept-up mass by SN 2000ft}

Our late time radio monitoring of SN 2000ft also allowed us to shed light on a number of relevant aspects for the supernova-CSM interaction, such as the mass-loss history of the progenitor star and the swept-up mass by the supernova shock, $\msw$. If we assume a free expansion for SN 2000ft at an average speed of $v_s = 10^4$\kms\ for the first $t \approx 2127$ days (the last epoch of our observations at 8.4 GHz), and a standard wind velocity, $v_w = 10$\kms, it follows that the 8.4 GHz emission is currently probing the wind of the progenitor star around $\tau = t\,v_s/v_w \approx $ 5820 yr prior to its explosion.  As found in the previous section, the CSM density at the corresponding circumstellar wind was $n_w \simeq 5.1\EE{3}$ cm$^{-3}$, which is in agreement with expectations for RSG supernova progenitors.

The swept-up mass by the supernova shock is 

\begin{equation}
\msw = 4\,\pi\,\int^{r_{sh}}_{r_0}\, \rho_w\,r^2\,\ud r 
     = \mdot \, \frac{r_{sh} - r_0}{v_w} \, \msun
\label{eq,msw}
\end{equation}

where we have assumed a standard progenitor wind ($s=2; \rho_w \propto
r^{-s}$), and $r_{sh}$, $r_0$, and $v_w$ are the current shock radius
of the supernova ($r_{sh} = v_{sh}\,t$), the radius of the supernova
at shock break-out, and the velocity of the stellar wind, respectively.
Since $r_{sh} \gg r_0$, then Equation \ref{eq,msw} can be rewritten as

\begin{equation}
\msw \simeq \mdot \,r_{sh}/v_w = 0.10 M_{-4}\,v_{sh1}\,v_{w1}^{-1}\,t_1\, 
{\rm \msun} 
\label{eq,msw-norm}
\end{equation}

where $v_{sh1}=v_{sh}/10^{4}\,$\kms, and $t_1$ is the time since the
SN explosion, in years.  Since we have $M_{-4}=0.49$, $v_{sh1}=1$, and
$t_1$=5.9, the swept-up mass by the  supernova shock of
SN 2000ft is then $\msw \simeq 0.29$\msun.

\subsection{The circumnuclear starburst in NGC 7469}
\label{ccsnrate}

The most relevant parameters of a starburst are its star formation rate ($SFR$) and its core-collapse supernova (CCSN) rate. The CCSN rate is especially relevant in the study of LIRGs, since all these supernovae emit in radio (albeit not all are so strong emitters that current radio interferometers --despite their high sensitivity-- are able to detect all of them).  It is also very important that the (constant) CCSN rate, \ccsnrate, can be related to $SFR$ as follows:

\begin{eqnarray}
\ccsnrate & = & \int_{m_{\rm SN}}^{m_u}\, \Phi (m)\,dm
           \nonumber\\
 & = & SFR\, \left( \frac{\alpha-2}{\alpha-1} \right) \left(
          \frac{m_{\rm SN}^{1-\alpha} - m_u^{1-\alpha}}{m_l^{2-\alpha} 
           - m_u^{2-\alpha}} \right)
\label{eq:ccsnrate}
\end{eqnarray}

where $SFR$ is the (constant) star formation rate in \msunyr, $m_l$ and $m_u$ are the lower and upper mass limits of the initial mass function (IMF, $\Phi \propto m^{-\alpha}$), and $m_{\rm SN}$ is the minimum mass of stars that yield supernovae, assumed to be 8 \msun (e.g., Smartt et al. 2009).  Following the formalism described in Colina \& P\'erez-Olea (1992), the observed far-infrared luminosity of a starburst, \lfir, can be used to constrain \ccsnrate\ and $SFR$.  For NGC 7469, which has L$_{\rm IR}$(8-1000$\mu$m) = 4.5$\times 10^{11} \Lsun$ (Sanders et al. 2003).  we obtain a constant \ccsnrate\ $\gsim$ 0.75 SN/yr, and $SFR \gsim 40$\msunyr, assuming two thirds of the above IR luminosity are emitted by its associated circumnuclear starburst (Genzel et al. 1995), a Salpeter IMF ($\alpha = 2.35$; Salpeter 1955) for $m_l = 1 \msun$ and $m_u = 120 \msun$, and that all of the ionizing photons go into dust heating.  If a fraction around 0.5 of all the photons go into dust heating, which is more realistic, then one obtains \ccsnrate\ $\lsim$ 1.5 SN/yr, and $SFR \lsim 80$\msunyr. Those values are in broad agreement with values previously obtained by Wilson et al. (1991), Colina \& P\'erez-Olea (1992), and Genzel et al. 1995), and also with expectations for core-collapse supernova rates based solely in measured far infrared luminosities, e.g., Mattila \& Meikle (2001), which found that \ccsnrate\ $\approx 2.7\EE{-12}$\,(\lfir/\Lsun)\,yr$^{-1}$, thus implying a CCSN rate for the circumnuclear starburst of NGC 7469 of $\approx$0.81 SN yr$^{-1}$, for L$_{\rm IR} = 3.0\times 10^{11}$.

We showed in Section \ref{results} that our 8.4 GHz observations indicate the explosion of only one very bright RSN (SN 2000ft; $L^{\rm peak}_\nu \approx 1.1\EE{28}$\ergshz) in the circumnuclear starburst of NGC 7469 during the period covered by our observations (7.8 years). This implies an upper limit for the average rate of very bright RSNe of $\lsim$0.13 SN/yr, which appears to be far off from the estimated (constant) CCSN rate for NGC 7469 of $\approx$(0.81--1.50) SN yr$^{-1}$, based on the FIR emission from its circumnuclear starburst.  To explain this disagreement, it has been usually advocated that not all CCSNe result in RSNe. Many CCSNe would then go undetected because they will not emit any radio emission.  However, this is a too simplistic view of CCSNe.  Since it is the interaction of the ejecta with circumstellar material that surrounds CCSN progenitors which gives rise to their radio emission, there is no reason to think this physical scenario is not valid for all CCSNe, all of which lose significant amounts of hydrogen-rich material in the last thousands of years of their lives through their stellar winds.  Therefore, all CCSNe are expected to generate some level of radio emission, albeit not all of them will result in very bright radio emitters, which coupled with the limited sensitivity of currently available radio interferometers will prevent the detection of faint RSNe.

Indeed, CCSNe display a broad range of radio luminosities, spanning several orders of magnitude, with essentially all detected supernovae showing peak luminosities from a few times $10^{25}$\ergshz\ (all of which would go undetected by our observations) up to a few times $10^{28}$\ergshz, or more (see e.g., Fig. 5 of Alberdi et al. 2006).  
If there was no background emission, RSNe brighter than about three times the off-source rms (rms=$1.5\EE{26}$\ergshz) could have been directly detected by our VLA-A observations.  In practice, direct detections of RSNe are limited by the background radio emission of the galaxy ($\approx 100 \mu$Jy), preventing us from detecting RSNe with $L_{\rm R} \lsim 6.0 \EE{26}$\ergshz).
Very bright, long-lived RSNe like SN 2000ft are expected to come essentially from Type IIL/IIn SNe.  Smartt et al. (2009) have found that these represent $\sim$6.5\% of all CCSNe (assuming a Salpeter IMF), while Type IIP/IIb SNe -the radio faint ones, with peak luminosities of $L_\nu^{\rm peak} \sim (5-20)\EE{25}$\ergshz- are much more numerous ($\sim$64.1\%).  (Type Ib/c SNe result often in bright radio events, but they fade away very quickly, typically in less than $\sim$100 days, so we do not take them into account, as chances of detecting them with our $\sim$1.5 yr monitoring of NGC 7469 are meager.)  Thus, whatever the CCSN rate, about 64.1\% of the exploding SNe are expected to become relatively radio faint. This would result in a RSN rate of $\approx (0.8-1.5) \times 0.36$=(0.29--0.54). Since Poisson statistics apply to the explosion of SNe, based on a single RSN discovery we cannot rule out that a starburst producing massive stars at a constant rate is taking place in NGC 7469.

Another possible explanation within a constant star forming scenario for NGC 7469 is that the radio luminosity function of CCSNe in (U)LIRGs is top-heavy, e.g., if the IMF results to be top-heavy, as suggested by, e.g., Klessen et al. (2007). In this scenario, we would be witnessing only the explosion of very massive stars, which yield also very bright radio emitters.  We must note, however, that if the radio luminosity function of CCSNe is bottom heavy (e.g., by mimicking a Salpeter IMF), then it is very unlikely that the only detected event in almost eight years would have been SN 2000ft, since similarly bright RSNe 
should have exploded during the 7.8 yrs covered by our VLA observing program.

Finally, we note that while a starburst producing stars at a constant rate cannot be excluded, \emph{Hubble Space Telescope} multi-wavelength (UV through NIR) imaging and $K$-band ground-based long-slit spectroscopy of NGC 7469 (D\'{\i}az-Santos et al. 2007) strongly suggest the existence of several localized starbursts in NGC 7469, which started at different times and in different locations of its circumnuclear ring.  According to the observations of D\'{\i}az-Santos et al. (2007) , the starburst in NGC 7469 consists of many massive, young, compact clusters characterized by masses between (1$-$10)$\EE{6}$\Msun\ and ages of several millions up to several tens of million years, and sizes of parsecs, which yield support to an scenario where many instantaneous bursts occur with a spread in ages.

\section{SN 2000ft and other very luminous radio supernovae in LIRGs}

SN 2000ft in NGC 7469 belongs, together with SN 2004ip in IRAS 18293-3413 (Mattila et al. 2007, P\'erez-Torres et al. 2007) and SN 2008cs in IRAS 17138-1017 (Kankare et al. 2008a, P\'erez-Torres et al. 2008, Kankare et al. 2008b), to the class of CCSNe that have been detected at both radio and optical/IR wavelengths in the nuclear regions of LIRGs. All of them have in common that are subject to a substantial extinction, and show a high radio luminosity. A question arises: Is this simply a selection effect?  The most luminous events at IR wavelengths are more easily detected by NIR searches that we are currently undertaking. The high IR luminosity can be expected to be powered by an interaction with a dense CSM, and the same scenario applies for high radio luminosity events.  Thus, significantly less luminous events would simply remain undetected by current VLA observations. Alternatively, it could be that such SNe are favoured in the nuclear regions of LIRGs, e.g. because of a top-heavy IMF, or a high metallicity, resulting in large mass loss rates for the SN progenitor stars, and therefore in a dense CSM. To get an unambiguous answer we need to keep searching for SNe in LIRGs at both IR and radio wavelengths so that we will have enough statistics in a few years from now.

\section{Summary}
\label{summary}

We have monitored the radio emission of the Luminous Infrared Galaxy (LIRG) NGC 7469 and its radio bright supernova SN 2000ft from 1998 until 2006, using the VLA at 8.4 GHz in A configuration.

Our 8.4 GHz VLA monitoring of SN 2000ft for almost six years ($t \approx$ 2100 days) has allowed us to show that its radio emission follows a rather steep decline ($\beta \simeq -2.02$; $S_\nu \propto t^\beta$), which is typical of supernovae exploding in ''normal'' environments.  This result implies that the late time radio emission of SN 2000ft is still being powered by its interaction with the presupernova stellar wind, rather than by interaction with the interstellar medium (ISM).  In fact, we find that the ram pressure of the presupernova wind is rather large $\rho_w\,v_w^{2} \approx 7.6\EE{-9}$\,${\rm dyn}\,{\rm cm}^{-2}$, even at a supernova age of $t \approx 2127$ days.  This value is about three times larger than the inferred ISM pressure in the starburst regions of M 82 at a similar distance.  Therefore, the pressure of the ISM where SN 2000ft exploded must be smaller than the ram pressure of the presupernova stellar wind, as otherwise we should have seen evidence of a flattening of the 8.4 GHz flux density time decay.  The swept-up mass by the supernova shock, after 2127 days of assumed free expansion in a steady, spherically symmetric wind ($s = 2; \rho \propto r^{-s}$), is $\msw \approx 0.29$\msun, and implies that the mass ejected in the explosion of SN 2000ft must have been significantly larger. At this SN age, the shock has reached a distance $r_{sh}\approx 0.06$ pc, and our 8.4 GHz VLA observations probe the interaction of the SN with material that was ejected by the progenitor star (via its stellar wind), about 5820 yr prior to its explosion.  This wind material has a particle density $n_w \simeq 5100$\cmmm, which is similar to expected values of $n_{\rm ISM}$ in the dusty, dense environments of starbursts.

We also searched in our radio images for recently exploded core-collapse supernovae (CCSNe).  Apart from SN 2000ft ($L_{\rm peak} \approx 1.1\EE{28}$\ergshz), we found no evidence for RSNe more luminous
than about ($L_{\rm peak} \approx 6.0\EE{26}$\ergshz), suggesting that no other Type IIn SN has exploded since 2000 in the circumnuclear starburst of NGC 7469.

\section*{Acknowledgments}
We thank an anonymous referee for a constructive report, that improved our manuscript.
This paper is based on observations with the Very Large Array (VLA) of the National Radio Astronomy Observatory (NRAO). NRAO is a facility of the National Science Foundation operated under cooperative agreement by Associated Universities Inc.  MAPT, AA, LC, and JMT acknowledge support by the Spanish Ministry of Education and Science (MEC) through grants AYA 2006-14986-C02-01 and AYA2008-06189-C03.  MAPT is a Ram\'on y Cajal Post Doctoral Research Fellow funded by the MEC and the Spanish Research Council (CSIC).  MAPT, AA, and JMT also acknowlegde support by the Consejer\'{\i}a de Innovaci\'on, Ciencia y Empresa of Junta de Andaluc\'{\i}a through grants FQM-1747 and TIC-126.  EK acknowledges support from the Finnish Academy of Science and Letters (Vilho, Yrj\"o, and Kalle V\"ais\"al\"a Foundation) and SM from the Academy of Finland (project 8120503).

\clearpage

\label{lastpage}
\end{document}